\documentclass[twocolumn]{article}
\usepackage{graphicx} % Required for inserting images
\usepackage{amsmath}
\usepackage{amsfonts}
\usepackage[a4paper,
            margin=2cm,
            includehead,
            includefoot]{geometry}
\usepackage{algorithm}
\usepackage{algpseudocode}
\usepackage{hyperref}
\usepackage{subcaption}
\usepackage[
style=numeric,
sorting=none,
backend=biber,
]{biblatex}
\graphicspath{figures/}

\usepackage{tikz}
\usetikzlibrary{positioning}
\usetikzlibrary {arrows.meta} 
\usetikzlibrary{calc}
\usetikzlibrary{backgrounds}
\usetikzlibrary {patterns,patterns.meta} 
\usetikzlibrary {decorations} 
\usetikzlibrary {decorations.pathreplacing} 
\usetikzlibrary {fit}
\usepackage{authblk}
\makeatletter
\def\@fnsymbol#1{}
\makeatother
\title{\texttt{ffdas}: volumetric ultrasound reconstruction at warp speed}
\author{
  Luuk Verhoef\thanks{BrainEcho Lab - Department of Neuroscience \& Neurosurgery, 
  Erasmus MC, Rotterdam, The Netherlands. 
  E-mail: \texttt{l.verhoef@erasmusmc.nl}, 
  \texttt{p.kruizinga@erasmusmc.nl}} \quad
  Pieter Kruizinga
}

\begin{document}

\maketitle
\begin{abstract}
Volumetric ultrafast ultrasound imaging demands reconstruction of images with millions of voxels thousands of times per second, creating computational challenges that limit both real-time feedback and easy offline analysis. Graphics processing units (GPUs) are well suited to this workload, yet we show that standard delay-and-sum implementations underutilize GPU resources through fragmented memory access patterns, even when sufficient computational capacity is available. Three optimization strategies address this: aligning memory access with GPU transfer granularity, halving memory traffic through mixed-precision storage, and exploiting spatial locality to utilize tensor core arithmetic. Together, these achieve kilohertz frame rates for $128^3$-voxel grids with 1024-element arrays, substantially outperforming existing implementations while maintaining image quality. This enables real-time volumetric imaging at scales previously restricted to offline processing, supporting applications such as intraoperative brain imaging and brain-computer interfaces where immediate feedback is essential. We release our implementation as part of the open-source \texttt{ffdas} library.
\end{abstract}

\section{Introduction}
Ultrafast ultrasound made it possible to image at frame rates which are orders of magnitude higher than conventional ultrasound \cite{tanter2014ultrafast}. This opened up a range of techniques that resolve fast and slow phenomena alike. For example: shear wave elastography~\cite{sandrin2002shear}, ultrafast Doppler~\cite{bercoff2011ultrafast}, pulse wave~\cite{kruizinga2014high} and electromechanical wave imaging~\cite{provost2011electromechanical}, functional ultrasound (fUSi)~\cite{mace2011functional}, and ultrasound localization microscopy (ULM)~\cite{errico2015ultrafast}.

These applications have so far been almost entirely two-dimensional. The natural next step is to do this in three dimensions. 4D ultrafast ultrasound brings volumetric imaging at high volume rate, capturing an entire region at once rather than a single plane, and this is what makes for example volumetric fUSi and ULM possible. It relies on hardware that has only recently become available: microbeamforming probes, fully populated matrix arrays with thousands of elements, and research scanners with matching receive channel counts.

Even with the hardware in place, reconstruction remains a bottleneck. The raw channel data must be turned into images, a step called beamforming, and for real-time imaging this has to keep pace with acquisition: each volume reconstructed before the next one arrives. At 4D scale the demands are steep, because several factors compound: volume rates in the kilohertz range, volumes of millions of voxels, broad transmit wavefronts that contribute to every voxel on every transmission, and arrays of several thousand elements. Most existing beamformers cannot keep up, so the data must instead be stored and reconstructed offline. This fails wherever real-time feedback is required, and even offline it slows the iteration that makes exploring large datasets productive.

Delay-and-sum is the standard reconstruction method for many applications. Due to its simplicity, it is robust, well understood, and readily parallelized~\cite{Perrot_2021}. Since each output voxel can be computed independently, it maps naturally onto the GPU's massively parallel architecture. At 4D scale the cost is substantial: reconstructing a single $128^3$-voxel volume from a 1024-element array over 16 transmit/receive events requires roughly $3.4 \times 10^{10}$ accumulation operations. 
Luckily GPUs can nowadays handle far higher computational throughput. Most naive delay-and-sum implementations, however, fail to utilize this throughput due to inefficient data access patterns.

Because each voxel sums contributions scattered across many channels, the access pattern is irregular, and standard implementations waste much of their available bandwidth moving data that is never used. The problem, in other words, is not how fast the GPU can compute but how efficiently it can be fed. A second obstacle is memory itself: precomputing and storing every delay would require on the order of $100$~GiB, which is impractical even on many current high-end systems. We therefore compute these delays in-kernel rather than storing them, as described in Section~\ref{sec:optimization-strategies}.

We address both with three optimization strategies that together enable real-time 4D reconstruction at previously unattainable scales. First, a data reordering step for more efficient memory access. Second, a mixed-precision data conversion to halve memory traffic. Third, exploitation of spatial locality in the image to enable dense matrix operations with tensor cores. Together these optimizations yield kilohertz frame rates for typical 4D data sizes. This brings real-time imaging to scales previously restricted to slow offline analysis, facilitating applications like 4D functional ultrasound where immediate feedback is essential~\cite{Verhoef_2025}. 

The remainder of this paper develops these optimizations in detail and evaluates their impact on representative 4D workloads. The optimizations described here are implemented in \texttt{ffdas}, an open-source CUDA library for delay-and-sum and related primitives in ultrasound, photoacoustics, and similar domains. Alongside the optimized delay-and-sum kernels, \texttt{ffdas}\footnote{\url{https://github.com/BrainEchoLab/ffdas}} provides GPU-accelerated rank truncation for clutter filtering, structured grid interpolation, and other primitives common in ultrafast processing pipelines, with Python and MATLAB bindings that integrate directly with established GPU array libraries. Researchers using \texttt{ffdas} in their work are kindly asked to cite this paper.

\label{sec:introduction}
% \cite{verhoef2026ffdas}.
% TODO: details on where to find library? For thesis, keep placeholder citation

\begin{figure*}[t]
\includegraphics[width=\textwidth]{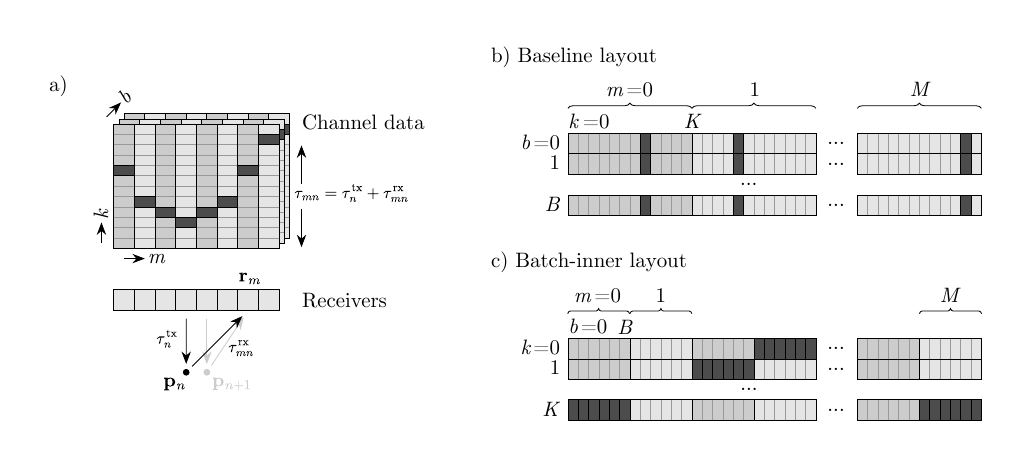}
\caption{\textbf{Delay-and-sum geometry and data layout.}
(a) Imaging geometry for voxel $\mathbf{p}_n$, showing transmit delay $\tau^\mathrm{tx}_{n}$
and receive delays $\tau^\mathrm{rx}_{mn}$ to receivers $\mathbf{r}_m$.
Required samples (dark) are highlighted in the channel data.
Neighboring voxel $\mathbf{p}_{n+1}$ has similar path lengths, causing nearby sample indices.
(b) Baseline memory layout with samples $k$ contiguous within each receiver $m$.
Samples required for $\mathbf{p}_n$ fall at irregular positions along $k$,
so threads processing neighboring voxels request nearby addresses within the same
memory sectors.
(c) Batch-inner layout with batch index $b$ contiguous within each $(m, k)$ combination.
Threads processing the same voxel for different batch items now access consecutive
addresses, aligning with GPU transfer granularity.
Observation dimension $q$ omitted for clarity.}
\label{fig:fig1}
\end{figure*}

\section{Delay-and-sum}
\label{sec:delay-and-sum}

The core idea behind delay-and-sum is to form an image by aligning and coherently summing signals from an array of receivers, optionally compounding over multiple observations (transmit and receive events). 
Let \(\mathbf{p}_n\in\mathbb{R}^3\) denote a voxel position, \(\{\mathbf{r}_m\in\mathbb{R}^3\}_{m=1}^M\) the receiver positions, and \(q\in\{1,\dots,Q\}\) index distinct observations. 
The discrete (real or complex baseband) signal from receiver \(m\) during observation \(q\) is \(y_{qm}[k]\) for \(k\in\{1,\dots,K\}\). 
Throughout this work, we write signal indices as subscripts for compactness (e.g., \(y_{qm}[k]\) for sample $k$ of receiver $m$ during observation $q$) and use bracket notation \(y[q,m,k]\) when the arrangement of dimensions in memory is relevant.

In ultrasound imaging, the time-of-flight between \(\mathbf{p}_n\) and \(\mathbf{r}_m\) decomposes into transmit and receive parts,
\begin{align}
\label{eq:tof}
\tau_{qmn} \;&=\; \tau^{\mathrm{tx}}_{qn}\;+\;\tau^{\mathrm{rx}}_{mn},\quad\tau^\mathrm{rx}_{mn}\;=\;\|\mathbf{p}_n-\mathbf{r}_m\|/c_0,
\end{align}
with nominal sound speed \(c_0\). 
The transmit term \(\tau_{qn}^{\mathrm{tx}}\) depends only on sequence geometry, such as plane or diverging waves\footnote{Plane wave with unit direction \(\hat{\mathbf{d}}_q\): \(\tau^{\mathrm{tx}}_{qn}=-\,\hat{\mathbf{d}}_q\!\cdot\!\mathbf{p}_n/c_0\). Diverging wave from virtual source \(\mathbf{v}_q\): \(\tau^{\mathrm{tx}}_{qn}=\|\mathbf{p}_n-\mathbf{v}_q\|/c_0\).}. 
Other imaging techniques, such as photoacoustic imaging, may omit the transmit term, but the geometric principle remains the same.

We further define a weight $w_{qmn} = w^{\text{tx}}_{qn} \cdot w^{\text{rx}}_{mn}$, decomposed into transmit and receive contributions. The transmit weight $w^{\text{tx}}_{qn}$ is the magnitude of the transmit wavefield at voxel $\mathbf{p}_n$ for observation $q$, and depends on the sequence design (aperture apodization, element delays, and any explicit field shaping) and is precomputed. The receive weight $w^{\text{rx}}_{mn}$ depends on the geometric relationship between voxel $\mathbf{p}_n$ and element $m$, each element having an origin and a direction, and is computed in-kernel alongside the receive delay.

The image value $\hat{x}$ at position \(\mathbf{p}_n\) is then obtained by evaluating and summing over all channel data \(y_{qm}[k]\) at the respective \(\tau_{qmn}\) (Figure~\ref{fig:fig1}A):
\begin{equation}
\label{eq:das}
% \hat{x}_n \;=\; \sum_{q=1}^{Q}\;\sum_{m=1}^{M} \; w_{qmn}\;\widetilde{y}_{qm}\!(\tau_{qmn}),
\hat{x}_n = \sum_{q=1}^{Q} \sum_{m=1}^{M} w_{qmn}\, y_{qm}[\tau_{qmn}],
\end{equation}
where \(y_{qm}[\tau]\) denotes temporal interpolation of the discrete channel data for non-integer \(\tau\).
% where \(\widetilde{y}_{qm}(\cdot)\) denotes the interpolated signal. 
In this work we assume linear interpolation between the two adjacent samples \(y_{qm}[k]\) and \(y_{qm}[k+1]\), with $k=\lfloor \tau\rfloor$. Alternative interpolation schemes follow the same principles.

An image is obtained by evaluating \eqref{eq:das} on a grid of \(N=N_x\times N_y\times N_z\) voxels. In this work, we target the large problem dimensions that are typically encountered in ultrafast 4D ultrasound imaging:
\begin{align}
\mathrm{(voxels)} \, N &= 128^3 \approx 2.1\times 10^6,\\
\mathrm{(receivers)} \, M &= 32^2 = 1024, \\
\mathrm{(observations)} \, Q &= 16.
\end{align}

These numbers pose a significant computational challenge: a single image entails \(N\times M\times Q\approx 3.4\times 10^{10}\) geometric time-of-flight calculations and sample interpolations. Precomputing and storing every \(\tau_{qmn}\) as 32-bit floats would then require
\[
N\times M\times Q\times 4\ \text{bytes} \;\approx\; 1.37\times 10^{11}\ \text{bytes} \;\approx\; 128~\text{GiB},
\]
which is impractical, even on many current high-end systems.
The sample count $K$ determines the size of the input data but does not affect the per-voxel cost, since each voxel requires only a single interpolation per receiver and observation regardless of $K$.

Moreover, real-time imaging applications require computing hundreds or thousands of images per second.
We therefore consider a batch of $B$ inputs, storing the channel data as an array $\mathbf{y} \in \mathbb{C}^{B \times Q \times M \times K}$ with samples contiguous along $k$ for each combination of batch item, observation, and receiver (Figure~\ref{fig:fig1}B).
Evaluating \eqref{eq:das} for each batch item yields outputs $\hat{x}_{bn}$ for $b \in \{1,\dots,B\}$.

Accelerating delay-and-sum starts by observing that the value for each voxel (and batch item) is independent and can be computed in parallel. 
In a naive implementation, however, the sample indices \(\tau_{qmn}\) lead to irregular and large-strided memory access in \(y[b,q,m,k]\), which is inefficient on GPUs and limits achievable throughput. 
In the following sections, we introduce the minimal GPU background and develop data layout, precision, and tiling strategies that increase memory efficiency and throughput, with profiling evidence for their effect on both speed and image quality.

\section{Optimizing for the GPU}
\label{sec:optimization-strategies}

To achieve real-time performance for volumetric delay-and-sum reconstruction while avoiding prohibitive memory storage requirements, we develop three optimization strategies that systematically address memory access inefficiencies while preserving the core accumulation logic from \eqref{eq:das}. We begin by establishing a baseline implementation and identifying its primary bottlenecks.

\subsection{Baseline implementation and performance bottlenecks}

GPUs achieve high throughput by distributing work across thousands of parallel threads organized into groups of 32 called \emph{warps}.
All threads in a warp execute the same instruction at each cycle, issuing their memory requests collectively.
For delay-and-sum, the most straightforward mapping assigns each thread the accumulation in \eqref{eq:das} for a single voxel.
We refer to this approach as our \textbf{baseline (BL)} implementation and show it in Algorithm~\ref{alg:baseline}.
Even in this baseline, we incorporate two straightforward optimizations.
First, we compute geometric parameters on-the-fly: transmit delays $\tau^{\mathrm{tx}}_{qn}$ and transmit weights $w^{\mathrm{tx}}_{qn}$ are precomputed per voxel, while receive delays and weights are calculated in-kernel from the receiver geometry.
This avoids constructing and storing the full, impractically large delay array in memory (as outlined in Section~\ref{sec:delay-and-sum}).
Second, each thread processes a small tile of $B_\mathrm{tile}$ consecutive batch items, reusing geometric calculations across them.
We choose $B_\mathrm{tile}=8$, but the optimal value will vary between systems.

\begin{algorithm}
\caption{Baseline delay-and-sum kernel (BL).}
\label{alg:baseline}
\begin{algorithmic}[1]
\Require $\mathbf{y} \in \mathbb{C}^{B \times Q \times M \times K}$, samples contiguous in $k$;
voxel positions $\{\mathbf{p}_n\}$; receiver positions $\{\mathbf{r}_m\}$;
transmit delays $\tau^{\text{tx}}_{qn}$; transmit weights $w^{\text{tx}}_{qn}$.
\Statex \textbf{Launch:} Grid over $(N, \lceil B/B_\text{tile} \rceil)$. Each thread computes one voxel $n$ for $B_\text{tile}$ batch items from offset $b_0$.
\State Initialize $\hat{x}_b \gets 0$ for $b = b_0, \dots, b_0 + B_\text{tile} - 1$
\For{$q = 1$ \textbf{to} $Q$}
    \For{$m = 1$ \textbf{to} $M$}
        \State $\tau^{\text{rx}}_{mn} \gets \|\mathbf{p}_n - \mathbf{r}_m\| / c_0$ \Comment{receive delay}
        \State $w^{\text{rx}}_{mn} \gets \mathrm{ReceiveWeight}(\mathbf{p}_n, \mathbf{r}_m)$ 
        \State $\tau \gets \tau^{\text{tx}}_{qn} + \tau^{\text{rx}}_{mn}$
        \State $w \gets w^{\text{tx}}_{qn} \cdot w^{\text{rx}}_{mn}$
        \State $(k, \alpha) \gets \mathrm{FractionalIndex} 
        (\tau)$ \Comment{$k = \lfloor \tau \rfloor$, $\alpha = \tau - k$}
        \For{$j = 0$ \textbf{to} $B_\text{tile} - 1$}
            \State $b \gets b_0 + j$
            \State $s_1 \gets \mathbf{y}[b, q, m, k]$; \quad $s_2 \gets \mathbf{y}[b, q, m, k{+}1]$
            \State $\hat{x}_b \gets \hat{x}_b + w \cdot \bigl( (1 - \alpha)\, s_1 + \alpha\, s_2 \bigr)$
        \EndFor
    \EndFor
\EndFor
\State \Return $\hat{x}_b$ for all $b$ in tile
\end{algorithmic}
\end{algorithm}

This approach benefits significantly from spatial locality in the delay-and-sum computation.
Neighboring voxels in regular grids have similar distances to each receiver (Figure~\ref{fig:fig1}A), causing threads in a warp to request nearly identical sample indices $k$ for any given $(q, m)$ pair.
As we will confirm through profiling in Section~\ref{sec:performance-evaluation}, this locality enables the GPU's caching mechanisms to serve most requests from fast L1 and L2 caches rather than DRAM, making the problem tractable despite its large size.

However, spatial locality alone does not solve the fundamental memory efficiency problem.
Because threads in a warp process neighboring voxels, they request nearly the same sample index $k$ from each receiver.
In the baseline data layout, where samples are contiguous along $k$, these requests map to the same or immediately adjacent memory addresses.
The GPU's memory system transfers data in fixed 32-byte sectors, so each request loads a full sector of consecutive time samples at that address, most of which no thread in the warp actually needs.
We refer to the fraction of transferred bytes that are actually used as \emph{payload efficiency}.

These misaligned access patterns generate many small, fragmented memory transactions that prevent optimal GPU performance. Even with high cache hit rates partially mitigating the problem, threads spend most execution time \emph{stalled}, waiting for memory operations rather than performing useful arithmetic. The key insight is that spatial locality provides the foundation for more efficient memory access, and the challenge lies in organizing computation to fully exploit this locality in a way that aligns with the GPU's architectural constraints.

To quantify these inefficiencies and track improvements, we monitor payload efficiency, request granularity (sectors per memory instruction), total DRAM traffic, and the distribution of execution time between memory stalls and arithmetic operations.

\subsection{Batch-inner tiling}

If, instead of assigning each thread its own voxel, multiple threads in a warp process the same voxel but for different batch items $b$, they would request the same sample index $k$ from each receiver.
Making $b$ the innermost (contiguous) dimension then places the values these threads need at consecutive addresses, matching the GPU's transfer granularity (Figure~\ref{fig:fig1}c).
We refer to this approach as \textbf{batch-inner (BI)} tiling.

We permute the input data from $B \times Q\times M\times K$ layout (with samples contiguous in $k$) to $K \times Q \times M \times B$ layout (with batch indices $b$ contiguous).
Each warp then processes a tile of $N_{\mathrm{warp}}$ neighboring voxels and $B_{\mathrm{warp}}$ batch items, for a total of $N_{\mathrm{warp}} \times B_{\mathrm{warp}}$ outputs.
We choose this product to be a multiple of the warp size (32) so that each thread computes the same number of outputs.

Algorithm~\ref{alg:batch-inner} shows the resulting kernel.
The warp is divided into $N_{\mathrm{warp}}$ groups of $G = 32/N_{\mathrm{warp}}$ threads, where each group processes one voxel for all $B_{\mathrm{warp}}$ batch items.
Within a group, thread $t \in \{0, \dots, G\!-\!1\}$ handles batch items $b = t, t\!+\!G, t\!+\!2G, \dots$, so that at each iteration the $G$ threads in the group access consecutive values of $b$ at the same sample location $(k, q, m)$.
Because $b$ is the contiguous memory dimension, these accesses map to adjacent addresses.

\begin{algorithm}[ht]
\caption{Batch-inner tiled delay-and-sum kernel (BI).}
\label{alg:batch-inner}
\small
\begin{algorithmic}[1]
\algrenewcommand\algorithmicensure{\textbf{Launch:}}
\Require $\mathbf{y} \in \mathbb{C}^{K \times Q \times M \times B}$, batch index $b$ contiguous;
  geometric inputs as in Algorithm~\ref{alg:baseline}.
\Ensure Grid over $(\lceil N/N_{\mathrm{warp}} \rceil,\; \lceil B/B_{\mathrm{warp}} \rceil)$.
  Groups of $G = 32/N_{\mathrm{warp}}$ threads share each voxel $n$;
  thread $t \in \{0,\dots,G\!-\!1\}$ processes $B_{\mathrm{warp}}/G$ batch items
  from offset $b_0$.
\State Initialize $\hat{x}_{b} \leftarrow 0$ for all assigned $b$
\For{$q = 1$ to $Q$}
    \For{$m = 1$ to $M$}
        \State $w \leftarrow w^{\mathrm{tx}}_{qn} \cdot w^{\mathrm{rx}}_{mn}$
        \State $\tau \leftarrow \tau^{\mathrm{tx}}_{qn} + \tau^{\mathrm{rx}}_{mn}$
        \State $(k,\, \alpha) \leftarrow \mathrm{FractionalIndex}(\tau)$
        \For{$j = 0$ to $B_{\mathrm{warp}}/G - 1$}
            \State $b \leftarrow b_0 + jG + t$
                \Comment{threads $t\!=\!0,\dots,G\!-\!1$ access consecutive $b$}
            \State $s_1 \leftarrow y[k,\, q,\, m,\, b]$;\quad
                   $s_2 \leftarrow y[k\!+\!1,\, q,\, m,\, b]$
            \State $\hat{x}_{b} \leftarrow \hat{x}_{b}
              + w \cdot \big((1\!-\!\alpha)\,s_1 + \alpha\,s_2\big)$
        \EndFor
    \EndFor
\EndFor
\State \Return $\hat{x}_{bn}$ for all $(b,\, n)$ in tile
\end{algorithmic}
\end{algorithm}

This organization enables several optimizations.
The memory system can \emph{coalesce} the group's individual requests into wide transactions that fully utilize each 32-byte sector, since threads access consecutive addresses at each iteration.
Because the batch dimension is contiguous in memory, each thread can issue \emph{vectorized loads} that fetch multiple values in a single instruction.
Finally, threads within a group that share the same voxel can distribute their geometric calculations through \emph{warp shuffles}, avoiding redundant computation of delays and weights.

The tile dimensions must balance competing factors: larger tiles provide more arithmetic work to hide memory latency but consume more register memory, potentially reducing \emph{occupancy} (the number of warps that can run simultaneously on each streaming multiprocessor).
In our implementation, we use $B_{\mathrm{warp}} = 64$ and $N_{\mathrm{warp}} = 2$, yielding groups of $G = 16$ threads that each process four batch items.
The best choice will depend on the specific hardware, and in practice should be determined by profiling different options.

This optimization requires an upfront permutation of input data, but this cost is generally negligible compared to the delay-and-sum runtime for moderate and large data sizes.
By design, payload efficiency should approach the full 32-byte sector size, and the GPU should generate fewer but wider memory requests.
We quantify these effects in Section~\ref{sec:performance-evaluation}.

\subsection{Mixed-precision storage}

We can improve memory access efficiency even further by reducing the total volume of transferred data. Most practical ultrasound acquisition systems produce samples with precision and dynamic range requirements well below those provided by 32-bit floating-point representation (FP32). We therefore store input samples in 16-bit floating-point (FP16) format, doubling the effective memory bandwidth by halving the storage requirement per sample. To maintain numerical accuracy in the output, samples are converted to 32-bit precision before interpolation and accumulation. We denote versions of our implementations with different precision with \textbf{-FP16} and \textbf{-FP32} suffixes.

The conversion cost is minimal as it can be combined with the data permutation to batch-inner layout. Precision reduction provides the greatest benefit when combined with efficient memory access patterns: while the baseline approach often transfers the same number of sectors regardless of element size (since requests typically span less than one sector), the batch-inner layout with its high payload efficiency directly translates reduced precision into proportionally less data transfer.

\subsection{Tensor core acceleration}

As memory access efficiency improves, floating-point arithmetic becomes an increasingly significant bottleneck.
Modern GPUs provide specialized \emph{tensor cores} that execute matrix multiply-accumulate operations with significantly higher throughput than standard scalar units, but require computations to be organized into small dense matrix tiles of specific dimensions.
General tensor core beamforming libraries have shown substantial performance gains across multiple domains \cite{Oostrum_2025}, but time-domain delay-and-sum does not naturally map to a dense matrix operation.
Because each voxel requires only two samples $(k, k+1)$ from each receiver, the computation is globally sparse.

Still, spatial locality (small spacing between voxels and receivers) makes it possible to repackage this sparse computation into dense tiles.
For a group of $N_{\mathrm{warp}}$ neighboring voxels, the required sample indices from a given receiver cluster tightly due to similar path lengths.
We exploit this by processing receivers in tiles of $M_{\mathrm{tile}}$ and iterating over the distinct sample positions that appear within each tile.
The value of $M_{\mathrm{tile}}$ is determined by the dimensions of the MMA instructions available on the target hardware, as receivers occupy the reduced dimension of the matrix product.

For each receiver $m$ in the tile, the warp computes delays for all $N_{\mathrm{warp}}$ voxels and collects the sorted set of unique sample indices $\mathcal{U}_m$ required across the group.
The iteration then proceeds over the maximum number of unique indices across all receivers in the tile $U = \max_m |\mathcal{U}_m|$.
For each unique index $i\in\mathcal{U}_m$, the warp assembles two matrices: a weight matrix $\mathbf{W} \in \mathbb{C}^{N_{\mathrm{warp}} \times M_{\mathrm{tile}}}$, where each entry captures the interpolation weight that voxel $n$ assigns to receiver $m$'s $i$-th sample (or zero if that index is irrelevant for the pair), and a sample matrix $\mathbf{Y} \in \mathbb{C}^{M_{\mathrm{tile}} \times B_{\mathrm{warp}}}$, containing the corresponding sample values across batch items.
The product $\mathbf{W}\mathbf{Y}$ yields the partial contributions for all voxels and batch items simultaneously, mapping directly to tensor core matrix multiply-accumulate (MMA) instructions.
Accumulating over all unique indices $i$, receiver tiles, and observations produces the final image values.
Algorithm~\ref{alg:tensor-core} shows the full kernel.

The tighter the sample indices cluster within a voxel tile, the smaller $U$ is and the fewer MMA instructions are needed per receiver tile.
Since MMA operations work on fixed-size tiles, throughput depends directly on this clustering, which in turn depends on the imaging geometry: voxels far from the array have similar path lengths to each receiver and share most sample indices, while voxels close to the array show more variation.
We refer to this method as the \textbf{tensor core (TC)} approach.

This transformation is designed to shift the execution bottleneck from scalar arithmetic throughput to the much higher tensor core throughput, while maintaining the memory efficiency gains from the batch-inner layout.
We quantify the resulting improvement in \emph{operational intensity} (useful computation per transferred byte) in Section~\ref{sec:performance-evaluation}.

\begin{algorithm}[ht]
\caption{Tensor core delay-and-sum kernel (TC).}
\label{alg:tensor-core}
\small
\begin{algorithmic}[1]
\algrenewcommand\algorithmicensure{\textbf{Launch:}}
\Require Same inputs and data layout as Algorithm~\ref{alg:batch-inner};
  receiver tile size $M_{\mathrm{tile}}$.
\Ensure Grid over $(\lceil N/N_{\mathrm{warp}} \rceil,\; \lceil B/B_{\mathrm{warp}} \rceil)$.
  Each warp computes $N_{\mathrm{warp}}$ voxels $\times$ $B_{\mathrm{warp}}$ batch items
  from offsets $n_0$, $b_0$.
\State Initialize $\mathbf{X} \in \mathbb{C}^{N_{\mathrm{warp}} \times B_{\mathrm{warp}}} \leftarrow \mathbf{0}$
  \Comment{accumulates $\hat{x}_{b,n}$}
\For{$q = 1$ to $Q$}
    \For{each tile of $M_{\mathrm{tile}}$ receivers from offset $m_0$}
        \State Compute $(k_{qnm},\, \alpha_{qnm},\, w_{qnm})$
          \hfill $\forall\, n \in \{1,\dots,N_{\mathrm{warp}}\}$,\;
          $m \in \{1,\dots,M_{\mathrm{tile}}\}$
        \State Collect sorted unique indices
          $\mathcal{U}_m = \mathrm{unique}\!\big(\{k_{nm},\, k_{nm}\!+\!1\}_{n=1}^{N_\mathrm{warp}}\big)$
          \hfill $\forall\, m$
        \State $U \leftarrow \max_m |\mathcal{U}_m|$
            \Comment{Maximum count of unique indices}
        \For{$i = 1$ to $U$}
            \State $\mathbf{W}\![n,m] \leftarrow
              \begin{cases}
                w_{nm}\,(1\!-\!\alpha_{nm}) & \text{if } \mathcal{U}_m[i] = k_{nm} \\
                w_{nm}\,\alpha_{nm}         & \text{if } \mathcal{U}_m[i] = k_{nm}\!+\!1 \\
                0                           & \text{otherwise}
              \end{cases}$
              \hfill $\forall\, n,m$
            \State $\mathbf{Y}\![m,b] \leftarrow
                \begin{cases}
               y\big[\mathcal{U}_m[i],\; q,\; m_0\!+\!m,\; b_0\!+\!b\big] & \text{if } i \leq |\mathcal{U}_m| \\
                0                           & \text{otherwise}
              \end{cases}$
              \hfill $\forall\, m,b \in \{1,\dots,B_{\mathrm{warp}}\}$

            % \State $\mathbf{Y}\![m,b] \leftarrow
            %   y\big[\mathcal{U}_m[i],\; q,\; m_0\!+\!m,\; b_0\!+\!b\big]$
            %   \hfill $\forall\, m,b \in \{1,\dots,B_{\mathrm{warp}}\}$
            \State $\mathbf{X} \leftarrow \mathbf{X}
              + \mathbf{W}\,\mathbf{Y}$
              \Comment{MMA: $\mathbb{C}^{N_{\mathrm{warp}} \times M_{\mathrm{tile}}}
                \cdot \mathbb{C}^{M_{\mathrm{tile}} \times B_{\mathrm{warp}}}$}
        \EndFor
    \EndFor
\EndFor
\State \Return $\mathbf{X}$ as $\hat{x}_{bn}$ for all $(b,\, n)$ in tile
\end{algorithmic}
\end{algorithm}

\section{Performance evaluation}

\label{sec:performance-evaluation}

We evaluate each optimization strategy systematically on representative ultrafast 4D ultrasound workloads, demonstrating how the design choices from Section~\ref{sec:optimization-strategies} translate to measurable performance improvements. Our evaluation combines detailed profiling of individual optimizations with end-to-end scaling analysis and comparison to existing implementations.

\subsection{Experimental setup}

\paragraph{Problem normalization.} 
For broad interpretability across imaging setups, spatial quantities are reported in wavelengths ($\lambda = c_0 / f_c$), and the IQ sampling rate as $s_\lambda$ samples per wavelength. Receiver pitch and image spacing are given in multiples of $\lambda$, and the sample count $K$ follows from $s_\lambda$ and the spatial extent of the image or volume.

\paragraph{Profiling methodology.} We distinguish kernel-level profiling, where we measure isolated GPU kernels to report payload efficiency, request granularity, DRAM traffic, and stall distribution, from end-to-end runtime measurements that include any required data permutation and conversion but exclude host-device transfers. Kernel-level analysis uses NVIDIA Nsight Compute \cite{nvidia_nsight_compute} to capture hardware performance counters, while end-to-end timing measures total reconstruction latency for practical applications.

\paragraph{Synthetic data generation.}
The reconstruction algorithms operate on arrays of IQ samples regardless of how
those samples were generated, so the performance measurements in this work are
independent of the simulation's physical fidelity.
For reproducibility, we briefly describe the simulation included in \texttt{ffdas}.
Scatterers are distributed along a trefoil knot centerline with small radial jitter.
Each scatterer acts as a point reflector, and the signal at each receiver is
computed from two-way propagation using the free-space Green's function
$G(\omega; \mathbf{r}, \mathbf{r}')
  = e^{-i\omega\|\mathbf{r}-\mathbf{r}'\|/c_0} / \|\mathbf{r} - \mathbf{r}'\|$.
The incident field at scatterer $\mathbf{s}_\ell$ for observation $q$ is
\begin{equation}
T_q(\omega; \mathbf{s}_\ell)
  = \sum_{m'=1}^{M} e^{-i\omega\delta_{qm'}}\,G(\omega;\, \mathbf{r}_{m'},\, \mathbf{s}_\ell),
\end{equation}
where $\delta_{qm'}$ are per-channel transmit time delays determined by the sequence geometry.
The channel data for receiver $m$ follow by summing the scattered responses:
\begin{equation}
Y_{qm}(\omega)
  = \sum_{\ell=1}^{L} G(\omega;\, \mathbf{s}_\ell,\, \mathbf{r}_m)\;
    T_q(\omega;\, \mathbf{s}_\ell).
\end{equation}
Delay-and-sum proceeds on time-domain IQ channel data, obtained by inverse FFT
of $Y_{qm}(\omega)$.

\paragraph{Evaluation parameters.} Unless noted otherwise, we use the problem dimensions shown in Table~\ref{:tab1}, chosen to represent typical ultrafast 4D ultrasound imaging scenarios. We compare baseline per-voxel mapping (BL), batch-inner tiling (BI), mixed-precision storage, and tensor core acceleration (TC), with both FP32 and FP16 input formats and FP32 accumulation throughout.
Scaling experiments vary one parameter at a time while keeping the others at the values listed in Table~\ref{:tab1}.

\begin{table}[ht]
\centering
\begin{tabular}{ll}
\hline
Parameter & Value \\
\hline
Batch size, $B$ & 128 \\
Sensor count, $M$ & $32 \times 32 = 1024$ \\
Observations, $Q$ & 1 \\
Samples, $K$ & 256 \\
Voxels, $N$ & $128^3 \approx 2.1 \times 10^6$ \\
Sensor pitch & $\lambda$ \\
Voxel spacing & $\lambda/2$ \\
GPU & RTX 4080 Super$^a$ \\
\hline
\end{tabular}
\caption{Evaluation and simulation parameters.\\$^a$The device's clock speed was locked during profiling for consistent measurements.}
\label{:tab1}
\end{table}

Note that our profiling uses a single observation ($Q = 1$).
Because delay-and-sum processes each observation independently, runtime scales linearly with $Q$, so achievable frame rates for compounding sequences with $Q$ observations are approximately $1/Q$ times the rates reported here.

\subsection{Overall performance impact}

\begin{figure*}[t]
\includegraphics[width=\textwidth]{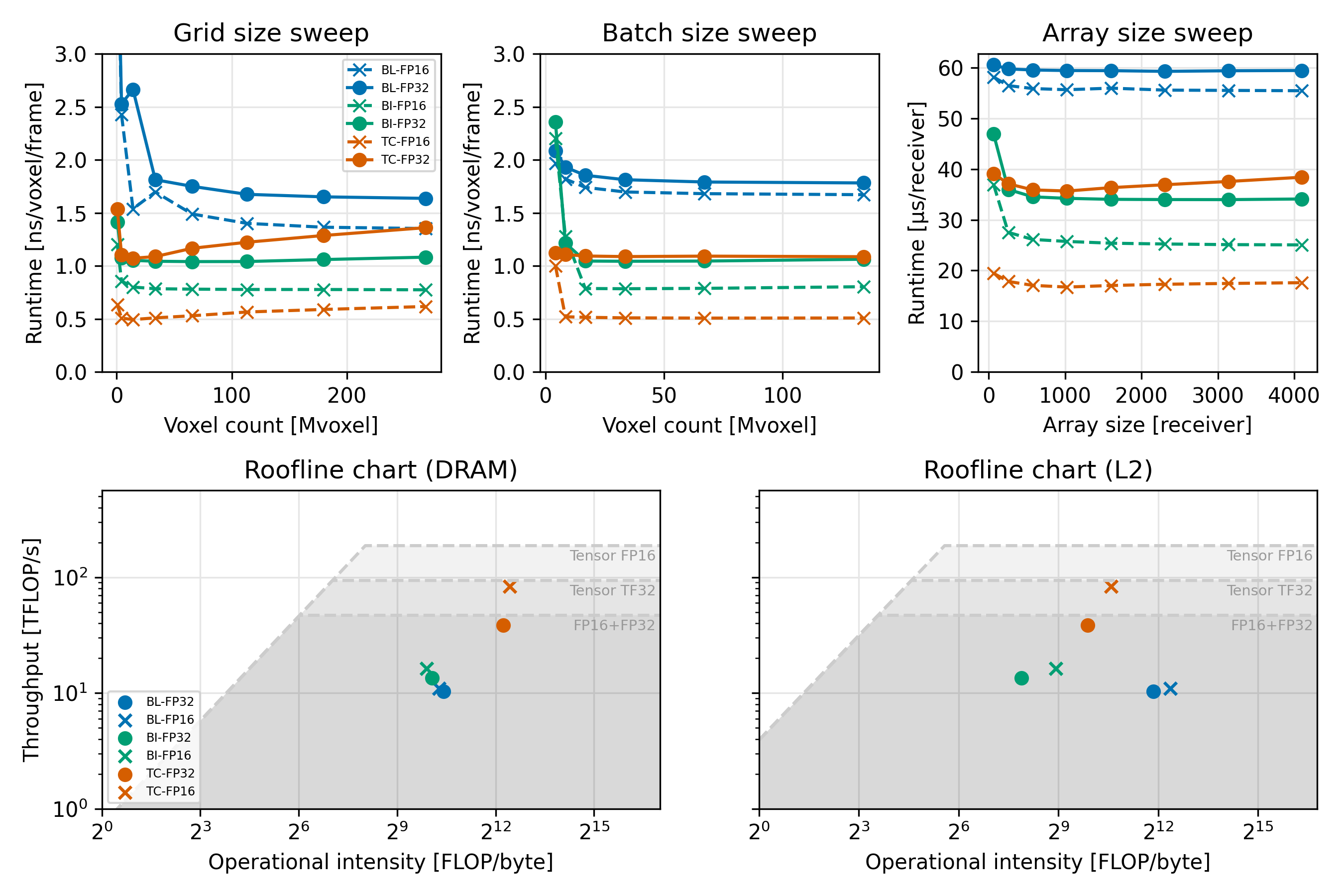}
\caption{
\textbf{Performance scaling and roofline analysis.} (A-C) Runtime scaling across grid size, batch size, and array size for baseline (BL), batch-inner (BI), and tensor core (TC) implementations, with FP16 and FP32 inputs. (D-E) DRAM and L2 roofline analysis showing operational intensity vs throughput.
}
\label{fig:fig2}
\end{figure*}
Figure~\ref{fig:fig2}A-C demonstrate the cumulative effect of our optimizations across representative problem dimensions. Scaling is near-linear once problems are large enough to saturate the GPU, with the BI implementation achieving a $\sim\!40\%$ speedup over baseline, and FP16 storage providing an additional $\sim\!20\%$ improvement where memory bandwidth remains limiting. The TC implementation further increases throughput in the compute-bound regime, primarily when combined with FP16 inputs, as tensor core units are optimized for mixed-precision processing.
These panels report relative scaling as individual parameters vary; absolute runtimes for the reference configuration in Table~\ref{:tab1} can be read from the comparison in Figure~\ref{fig:fig5}.

The tensor core efficiency depends on how tightly sample indices cluster within each warp tile.
Far from the array, the path length to a given receiver varies minimally across neighboring voxels, so most voxels in a tile share the same sample index and the weight matrix is small.
Close to the array, the path lengths to this receiver diverge more steeply between neighboring voxels.
This increases the number of unique sample indices per tile and requires more MMA instructions per receiver, reducing the throughput advantage of the TC implementation (visible in Figure~\ref{fig:fig2}A as the grid extent increases beyond the array).

A roofline analysis relates each implementation's achieved arithmetic throughput to its operational intensity (arithmetic operations per byte transferred), relative to the hardware's peak bandwidth and compute limits.
Implementations to the left of the ridge point are bandwidth-limited; those to the right are compute-limited.
In either regime, the vertical distance to the roofline indicates how much of these resources goes unused.
Figure~\ref{fig:fig2}D-E shows this analysis at two levels of the memory hierarchy.
On both rooflines, all implementations sit at high operational intensity, well into the compute-bound region, indicating that bandwidth capacity is not the bottleneck at either level.
The BL implementation sits furthest to the right on the L2 roofline, meaning that each value loaded from L2 experiences the highest reuse.
Yet despite this apparently favorable position, BL achieves the lowest throughput, far below the compute ceiling.
As the next section will show, BL's high reuse stems from many small, fragmented L1 requests that individually hit the cache but collectively stall threads waiting for the memory system to process them.
The BI and TC implementations reuse each L2 value less (lower operational intensity) but structure their memory access so that arithmetic can proceed in parallel with transfers, yielding higher throughput overall.
The TC implementations additionally benefit from the higher compute ceiling of the tensor core units.
FP16 storage increases operational intensity across all implementations by halving the transferred bytes, though the throughput gain is largest for the implementations whose regular access patterns already make full use of each transfer.

\subsection{Memory access efficiency}

\begin{figure*}[ht]
\includegraphics[width=\textwidth]{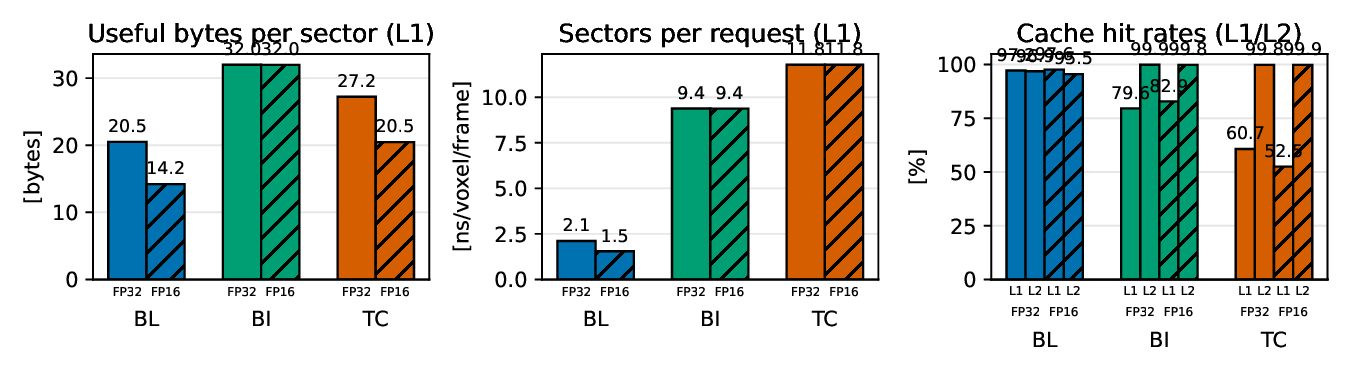}
\caption{
\textbf{Memory access efficiency.} (A) Payload efficiency increases from 40\% to 100\% with batch-inner layout. (B) Request consolidation reduces memory fragmentation, with wider transactions improving bandwidth utilization. (C) Cache hit rates across L1 and L2 levels, showing trade-offs between hit rate and request granularity.
}
\label{fig:fig3}
\end{figure*}
The performance improvements stem primarily from more efficient memory bandwidth utilization. Figure~\ref{fig:fig3}A quantifies the core mechanism: payload efficiency increases from approximately 40\% (13 useful bytes per 32-byte sector) for BL to 100\% (32 useful bytes per sector) for the batch-inner layout with warp-level tiling. This improvement reflects proper alignment of memory access patterns with GPU transfer granularity.

The batch-inner layout also allows the hardware to consolidate the warp's memory requests into fewer, wider transactions (Figure~\ref{fig:fig3}B) as they follow regularly spaced patterns.
In the BL implementation, threads request the same few sample values, so their requests collide at the same addresses rather than spanning a contiguous range, producing many narrow transactions (1--2 sectors per warp instruction).
In the BI implementation, threads access different batch items at consecutive addresses, so the hardware combines their requests into wide contiguous transactions (9.4 sectors per instruction on average).
The TC implementation achieves even wider requests (12 sectors) because it loads complete sample tiles where each thread fetches a distinct value, with data reuse occurring during the matrix multiply-accumulate operations rather than at the memory access level.

Mixed-precision storage amplifies these benefits when combined with efficient access patterns. FP16 input storage halves bandwidth requirements for the optimized implementations while providing minimal benefit to the BL implementation, whose poor payload efficiency wastes most transferred bytes regardless of element size.

Cache behavior provides important context for these results. 
The BL implementation's high L1 and L2 hit rates (Figure~\ref{fig:fig3}C) partially mitigate its poor memory access pattern, as most requests are served from fast caches rather than slow DRAM. 
However, the many small, fragmented requests still introduce substantial overhead. 
The optimized approaches achieve lower hit rates but generate far fewer total requests, so the cumulative per-request overhead is reduced and compensates for the lower fraction of cache hits.

\subsection{Execution behavior analysis}

\begin{figure}[ht]
\centering
\includegraphics[width=0.5\textwidth]{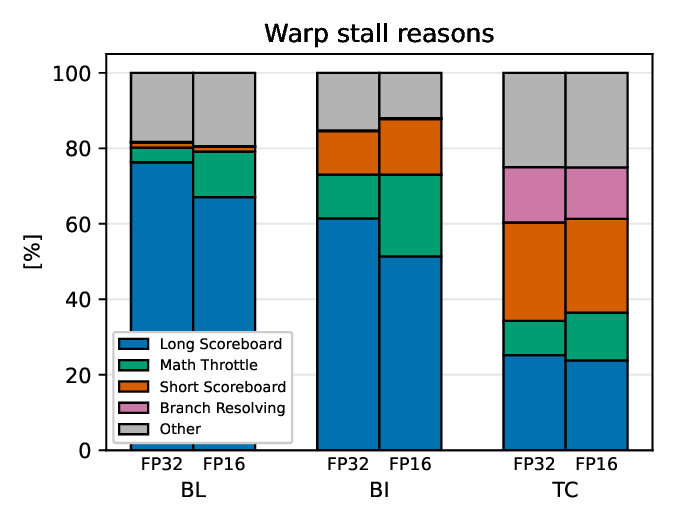}
\caption{
\textbf{Warp stall breakdown.} Distribution of warp stall reasons across implementations, showing the shift from memory-bound (Long Scoreboard) to more balanced execution with increased arithmetic pipeline utilization (Math Throttle) in optimized variants.
}
\label{fig:fig4}
\end{figure}

The variations in memory access patterns shift the bottlenecks that warps encounter during execution. Figure~\ref{fig:fig4} shows the normalized distribution of warp stall reasons across our implementations. The BL implementation spends most execution time in memory latency stalls (Long Scoreboard), indicating threads waiting for data transfers. While Long Scoreboard remains the dominant stall reason throughout, the BI implementation shows increased arithmetic pipeline pressure (Math Throttle), reflecting better balance between memory and compute utilization.

The TC implementation introduces additional complexity. 
Branch resolution increases due to the dynamic matrix assembly, introducing additional control flow and coordination between threads. 
Short Scoreboard stalls also increase significantly, reflecting implementation-specific factors such as shared memory usage and warp-level synchronization required for efficient matrix tile assembly. 
Despite these overheads, overall performance improves due to the higher arithmetic throughput from specialized matrix units.

These stall patterns confirm that our optimizations successfully reduce memory bottlenecks while shifting execution toward compute-intensive regimes, though memory latency remains a significant factor even in optimized implementations.

\subsection{External library comparison}

To demonstrate the practical effects of our optimizations, we compare our implementation against two established delay-and-sum libraries used in the ultrasound research community: \texttt{mach} (dedicated CUDA implementation) and \texttt{vbeam} (based on the JAX framework) \cite{Guan_2025, Kvalevag_2023, Bradbury_2018}. 
The comparison focuses purely on computational throughput, acknowledging that different tools serve different purposes. 
For example, \texttt{vbeam} prioritizes differentiability over raw throughput, enabling gradient-based optimization of the reconstruction pipeline, a fundamentally different and valuable design goal.
Our implementation targets the specific use case where reconstruction speed determines whether real-time visualization and further off-line data exploration is possible.  
We use our tensor core implementation with FP16 inputs (TC-FP16) as the representative method here, but the other options can be derived accordingly from the rest of our profiling results.
All libraries were configured to use the same reconstruction settings, as per Table~\ref{:tab1}.

\begin{figure}[ht]
\centering
\includegraphics[width=0.5\textwidth]{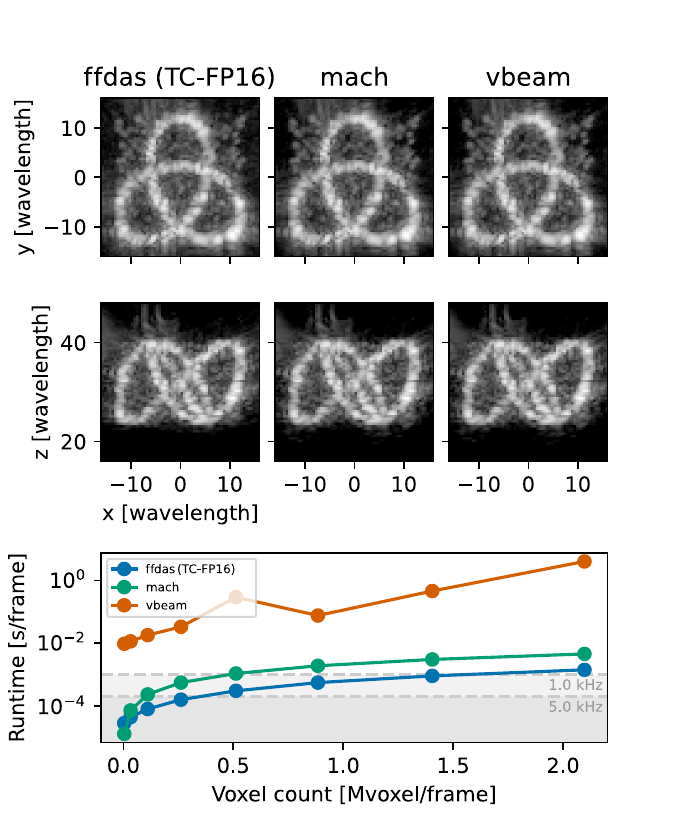}
\caption{
\textbf{External library performance comparison.} Runtime comparison with established delay-and-sum implementations (\texttt{mach}, \texttt{vbeam}) showing substantial performance improvements while maintaining visual quality under matched reconstruction settings. Each image is normalized individually and displayed with 32 dB dynamic range.
}
\label{fig:fig5}
\end{figure}

Figure~\ref{fig:fig5} demonstrates that existing implementations fall short of real-time requirements for large-scale 4D imaging. 
Under identical reconstruction settings, our optimized kernel achieves frame rates significantly higher than current alternatives for representative problem sizes ($3{-}4\times$ compared to \texttt{mach}, and $100{-}1000\times$ compared to \texttt{vbeam}). 
This speed-up enables imaging at higher spatial resolution, faster temporal sampling, or larger field-of-view without sacrificing real-time feedback.
Visual quality remains virtually identical across methods, confirming that our optimizations improve speed without compromising reconstruction fidelity.

\section{Conclusions}
\label{sec:conclusions}

Volumetric reconstruction at ultrafast rates is feasible on a consumer GPU. Our implementation reaches kilohertz frame rates for $128^3$-voxel volumes from 1024-element arrays. This brings 4D imaging to real-time scales that previously required offline processing, and reduces offline reconstruction times enough to make iterative exploration of large datasets practical.

Our profiling shows that the GPU has the capacity for this workload, but standard delay-and-sum implementations cannot extract it. Memory accesses fragment into many small, narrow transactions, and threads spend most of their time waiting for those transactions to complete rather than computing. The clearest expression of this is the baseline implementation, which sits deep in the compute-bound region of the roofline yet delivers low throughput. The bottleneck is not how fast the GPU can compute, but how efficiently it can be fed.

The three optimizations address this underutilization through complementary mechanisms. Batch-inner tiling permutes the input data so that threads in a warp access consecutive memory addresses, raising payload efficiency from 40\% to 100\% and consolidating many small requests into few wide ones.
Mixed-precision storage halves the data volume per sample, which, combined with the efficient access pattern, halves memory traffic and frees bandwidth for the optimized kernels to exploit. Tensor core acceleration exploits a different property of the delay-and-sum geometry: because neighboring voxels require nearly identical samples from each receiver, the globally sparse computation can be repacked into locally dense matrix operations that shift execution onto the GPU's highest-throughput arithmetic units.

Which kernel to use depends on the application. The baseline implementation requires no data permutation and uses less memory, making it suitable when GPU memory is scarce or when integration simplicity is prioritized over throughput.
The batch-inner layout with mixed-precision storage provides consistent improvement across all tested configurations and is the recommended default whenever a batch of volumes is reconstructed together, as in ultrafast Doppler, where it delivers consistent gains across the configurations we tested. The tensor core approach offers the highest throughput for dense voxel grids, where spatial locality ensures tight clustering of sample indices within each warp tile. Its advantage is largest for voxels far from the array and diminishes as path length variation across the tile increases, so the effective speedup depends on the imaging geometry.

Several directions could extend this work. First, emerging lower-precision formats such as 8-bit floating-point could further reduce memory traffic and unlock higher tensor core throughput on hardware that supports them, though the effect on image quality needs careful evaluation. Second, many transmit sequences concentrate energy in a limited part of the field of view, leaving most transmit weights near zero for a given voxel. Storing a sparse list of relevant observation indices per voxel would let the kernel skip these negligible contributions, reducing computation for spatially selective sequences without altering the output. Finally, extending the profiling methodology to other GPU architectures would clarify which of the bottlenecks we identify are intrinsic to delay-and-sum at this scale and which are specific to our reference hardware.

Beyond raw throughput, computing delays on-the-fly rather than precomputing and storing billions of parameters makes large-scale volumetric reconstruction practical on common workstations without requiring specialized high-memory hardware. The optimizations described here are implemented in \texttt{ffdas}, an open-source CUDA library for delay-and-sum and related primitives in ultrasound, photoacoustics, and similar domains. Alongside the optimized kernels, \texttt{ffdas} provides GPU-accelerated rank truncation for clutter filtering, structured grid interpolation, and other primitives common in ultrafast processing pipelines, with Python and MATLAB bindings that integrate directly with established GPU array libraries. Researchers using \texttt{ffdas} in their work are kindly asked to cite this paper.

\medskip

\printbibliography

\end{document}